\documentclass[twocolumn,,amsmath,amssymb]{revtex4}
\usepackage{graphicx}
\usepackage{dcolumn}
\usepackage{bm}

\newcommand{\be}{\begin{equation}}
\newcommand{\ee}{\end{equation}}
\newcommand{\bfig}{\begin{figure}}
\newcommand{\efig}{\end{figure}}

\begin{document}

\title{The magnetic phase diagram of underdoped YBa$_2$Cu$_3$O$_y$ inferred from torque magnetization and thermal conductivity}



\author{Fan Yu$^1$}
\author{Max Hirschberger$^2$}
\author{Toshinao Loew$^3$}
\author{Gang Li$^{1}$}
\author{Benjamin J. Lawson$^{1}$}
\author{Tomoya Asaba$^{1}$}
\author{J. B. Kemper$^{4}$}
\author{Tian Liang$^2$}
\author{Juan Porras$^3$}
\author{G. S. Boebinger$^4$}
\author{J. Singleton$^5$}
\author{B. Keimer$^3$}
\author{Lu Li$^{1}$}
\author{N. P. Ong$^{2,*}$}
\affiliation{
$^1${Department of Physics, University of Michigan, Ann Arbor, MI}\\
$^2${Department of Physics, Princeton University, Princeton, NJ 08544}\\
$^3${Max Planck Institute for Solid State Research, 70569 Stuttgart, Germany}\\
$^4${Dept. of Physics and National High Magnetic Field Laboratory, Florida State University, FL}\\
$^5${National High Magnetic Field Lab., Los Alamos National Lab., NM}
} 

\date{\today}      
\begin{abstract}
Strong evidence for charge-density correlation in the underdoped phase of the cuprate YBa$_2$Cu$_3$O$_y$ was obtained by nuclear magnetic resonance (NMR) and resonant x-ray scattering. The fluctuations were found to be enhanced in strong magnetic fields. Recently, 3D (three dimensional) charge-density wave (CDW) formation with long-range order (LRO) was observed by x-ray diffraction in $H>$ 15 T. To elucidate how the CDW transition impacts the pair condensate, we have used torque magnetization to 45 T and thermal conductivity $\kappa_{xx}$ to construct the magnetic phase diagram in untwinned crystals with hole density $p$ = 0.11. We show that the 3D CDW transitions appear as sharp features in the susceptibility and $\kappa_{xx}$ at the fields $H_K$ and $H_p$, which define phase boundaries in agreement with spectroscopic techniques. From measurements of the melting field $H_m(T)$ of the vortex solid, we obtain evidence for two vortex solid states below 8 K. At 0.5 K, the pair condensate appears to adjust to the 3D CDW by a sharp transition at 24 T between two vortex solids with very different shear moduli. At even higher $H$ (41 T) the second vortex solid melts to a vortex liquid which survives to fields well above 41 T. de Haas-van Alphen oscillations appear at fields 24-28 T, below the lower bound for the upper critical field $H_{c2}$. 
\end{abstract}
 
\maketitle



\section{Introduction}
The existence of static charge order in underdoped YBa$_2$Cu$_3$O$_y$ (YBCO) in an intense magnetic field was reported by Wu \emph{et al.} using nuclear magnetic resonance (NMR)~\cite{Julien}. Subsequently, Ghiringhelli \emph{et al.} ~\cite{Keimer2012} uncovered temperature-dependent charge density wave (CDW) correlations in zero magnetic field by resonant x-ray scattering (RXS). The RXS signal onsets near 180 K and peaks at the superconducting critical temperature $T_c$ before falling. Several groups~\cite{Chang2012,Blanco,Blackburn} recently showed that the RXS intensity is enhanced in finite field $H$. At low temperature $T$, several experiments have uncovered field-induced transitions. These include a transition at 18 T from ultrasonic measurements~\cite{LeBoeuf}, features in the high-field thermal conductivity~\cite{Grissonnanche}, and the onset of line splitting in the NMR spectra starting at the charge-ordering field $H_{ch}$ = 10-15 T and saturating at around 18 - 20 T (depending on the hole density $p$)~\cite{TaoWu2013}. Recent X-ray diffraction experiments~\cite{Gerber,Chang16,Tacon} show that the transition to a three-dimensional (3D) CDW with long-range order (LRO) onsets near $H$ = 15 T.

The field-induced CDW state raises several intriguing questions regarding its relation to superconductivity. Does CDW formation with LRO suppress the superconducting condensate? Where is the true upper critical field $H_{c2}$? How does it affect the stability of the vortex solid? While NMR and x-ray diffraction are incisive probes of charge modulation and CDW formation, they do not couple directly to the fundamental excitations of the pair condensate (quasiparticles and vortices), so they are less sensitive to the pairing correlations which reflect superconductivity. By contrast, the diamagnetic magnetization ${\bf M}_d$ (which dominates the observed magnetization ${\bf M}_{obs}$ below the critical temperature $T_c$) couples directly to the vortex excitations because it measures the current-current correlation. In the magnetic phase diagram, hysteretic behavior of the ${\bf M}_d$ curves readily identifies the stability region of the vortex solid. Above the melting field of the solid $H_m$, ${\bf M}_d$ also identifies the vortex liquid which displays a unique reversible diamagnetism. Surprisingly, we find that the magnetic susceptibility $\chi_d$ also detects the onset of charge ordering as a weak peak at the field $H_K$. Both $H_K$ and a higher cross-over field $H_p$ are apparent in the thermal conductivity. We compare $H_K$ and $H_p$ with field-scales reported from NMR and ultrasonic experiments. Taken together, these features fill out the magnetic phase diagram, and relate the charge-ordering fields to the phase boundaries in the vortex system.

\begin{figure}[ht]
\includegraphics[width=8 cm]{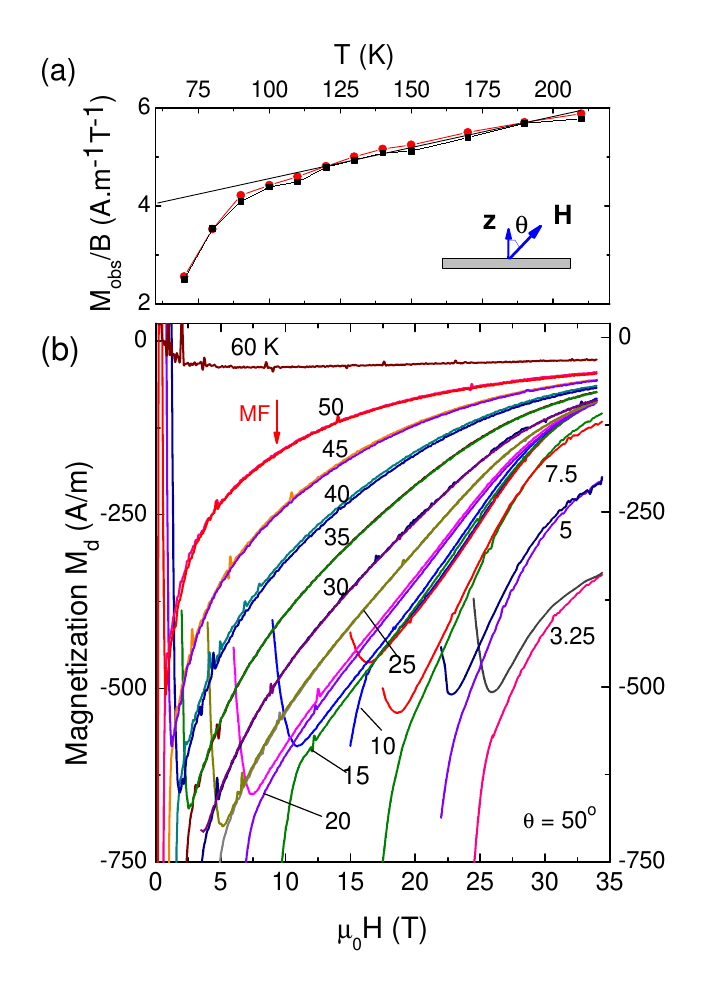}
\caption{\label{figMH} (color online) 
Magnetization in ortho-II YBa$_2$Cu$_3$O$_y$ ($y$ = 6.55, $T_c$ = 61 K, $p$ = 0.11) measured by torque magnetometry in dc field. Panel (a) shows the observed susceptibility $M_{obs}/B$ from $T$ = 80 to 210 K measured at 8 T (black circles) and 11 T (red) in Sample 1 ($B=\mu_0H$, with $\mu_0$ the permeability). The straight line is the Van Vleck  anisotropy background $\Delta\chi^v = a(T+T_0)$ with $a$ = 1.18$\times 10^{-2}$ A(mTK)$^{-1}$ and $T_0$ = 305 K. The diamagnetic term $M_d$ onsets near 110 K. The inset defines the tilt angle $\theta$ between $\bf H$ and the normal $\bf\hat{z}\parallel \hat{c}$. Panel (b) plots $M_d(T,H) = M_{obs}(T,H)-a(T+T_0)H$ vs. the magnetic field $\mu_0H$ measured at selected $T$ in Sample 2 with $\theta = 50^{\circ}$. Splitting of the sweep-up from the sweep-down curves occurs at the melting field $H_m(T)$. At each $T$, the sample is in the vortex liquid state for $H>H_m(T)$. Diamagnetism persists to fields significantly above 32 T. For the curve at 50 K, the red arrow (labelled ``MF'') marks the $H_{c2,MF}/\cos\theta$ using the mean-field value suggested for $H_{c2}$~\cite{Grissonnanche}. 
}
\end{figure}

We have measured by high-resolution torque magnetometry the magnetization and thermal conductivity $\kappa_{xx}$ in high-purity, detwinned ortho-II crystals of YBa$_2$Cu$_3$O$_y$ ($y$ = 6.55, $T_c$ = 61 K, $p$ = 0.11) in a dc field $H$ up to 45 T. Sample 1 has a large volume to facilitate accurate measurements at $T$ up to 210 K. Sample 2 and Sample 3 were picked to have smaller volumes (0.184 mm$^3$ and 0.344 mm$^3$, respectively) to avoid overloading the cantilever in the vortex solid regime below $T_c$ (see Ref. \cite{Blanco} for details on the crystals). Samples 4 and 5 were used for the $\kappa_{xx}$ measurements. From 120 to 210 K (Fig. \ref{figMH}a), the torque magnetization $M_{obs}$ arises predominantly from the anisotropy $\Delta\chi^v=\chi^v_c-\chi^v_a$ of the paramagnetic Van Vleck  term ($\chi^v_i$ is the susceptibility along axis $i$).~\cite{Wang2005,Li2010} $\Delta\chi^v(T)$ has a weak $T$ dependence $\Delta\chi^v= a(T+T_0)$, with $T_0$ = 305 K (see SI Appendix, Sec. S1). At 110 K, we start to resolve a negative contribution caused by fluctuating diamagnetism. As $T\to T_c$, the diamagnetic component $M_d$ grows rapidly in magnitude. In Sample 2, we focussed on measurements of $M_{obs}$ to 34 T at selected $T$ between 10 and 80 K. In non-superconducting LSCO ($x=0.050$), the Van Vleck  term retains the form $a(T+T_0)$ down to 10 K~\cite{Wang2005}. Hence to isolate $M_d$, we subtract this term from $M_{obs}$, i.e. $M_d(T,H) = M_{obs}(T,H) - a(T+T_0)H$.

\section{Diamagnetism in the vortex liquid state}
Figure \ref{figMH}b displays the curves of $M_d$ vs. $H$ measured at temperatures from 60 K down to 3.25 K in fields up to 34 T with tilt angle $\theta = 50^\circ$. At all $T<$ 60 K, $M_d$ is negative (diamagnetic) but increases rapidly with $H$. Throughout most of the field scale shown, the curves are \emph{reversible} (sweep-up and sweep-down traces coincide). The reversible, strongly $T$-dependent diamagnetic response is the hallmark of the vortex liquid. Throughout the vortex liquid regime, $|M_d|$ is far too large to be attributed to Gaussian fluctuations (100 A/m corresponds to 1.8$\times 10^{-3}$ Bohr magneton per unit cell). The diamagnetic signal reflects the array of small, diamagnetic supercurrent loops occupying the interstitial space between vortex cores. At each $T$, bifurcation of the sweep-up and -down curves occurs at the melting field $H_m(T)$ which separates the vortex solid from the liquid (at the scale shown this first becomes apparent at 45 K at $\sim$1.5 T). As $T$ decreases, $H_m(T)$ rises rapidly, reaching 25 T at 3.25 K. (As sketched in Fig. \ref{figMH}a, $\bf H$ is aligned at an angle $\theta$ to the $c$-axis of the crystal. We define ${\bf H\cdot\hat{c}} = H_z$.) 

\begin{figure}[ht]
\includegraphics[width=8 cm]{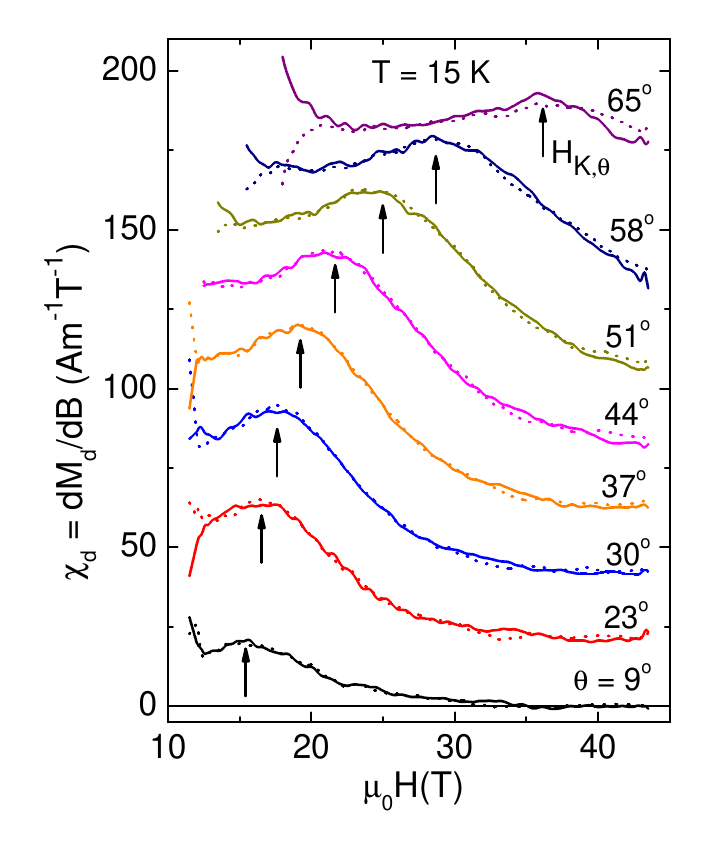}
\caption{\label{figChiAngle} (color online) 
The differential magnetic susceptibility $\chi_d \equiv \frac{dM_d}{dB}$ of YBCO (Sample 3) is plotted against the magnetic field $H$ at selected tilt angles $\theta$ at $T$ = 15 K. The curves have been shifted vertically for clarity. At each $\theta$, the solid (dashed) curve was recorded while the field $H$ was slowly swept up (down). The bifurcation of the curves at low fields defines the melting field $H_m$ of the vortex solid to the liquid state. In the liquid state, the $\chi_d$-$H$ curves display a broad peak which defines the field $H_{K,\theta}$ (arrows).}
 \end{figure}

\section{Peak in the differential susceptibility}
\begin{figure}[ht]
\includegraphics[width=7 cm]{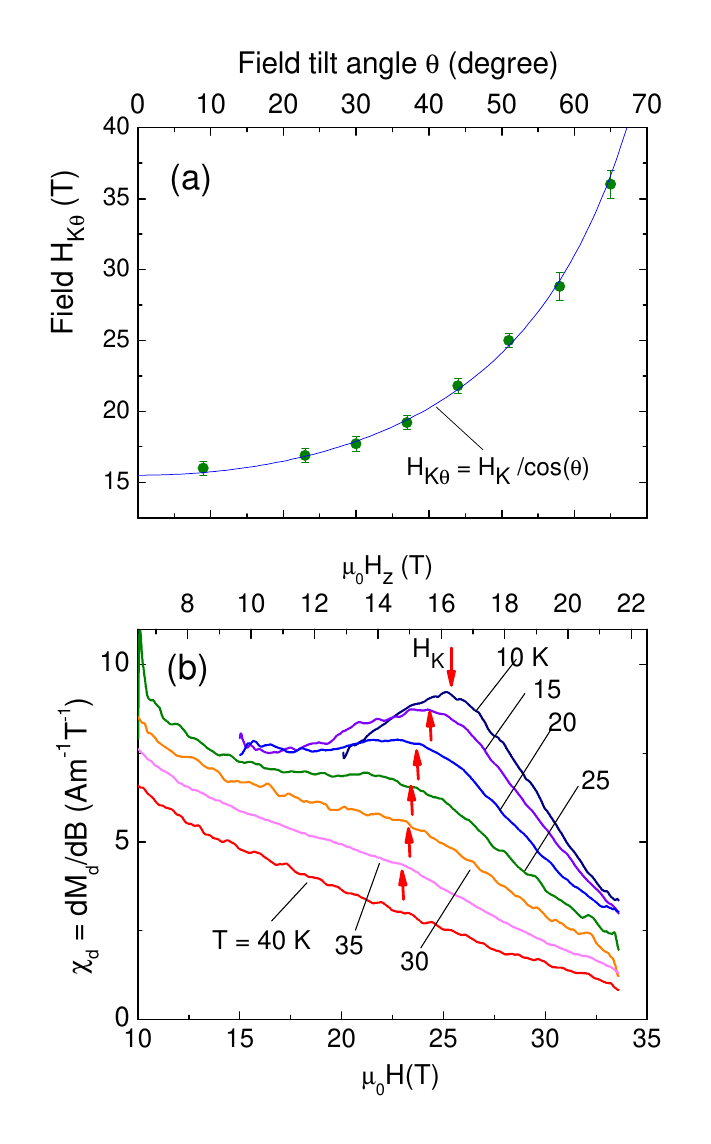}
\caption{\label{figAngle} (Panel a) The angular dependence of the kink field $H_{K,\theta}$ at 15 K in Sample 3 identified by the arrows in Fig. \ref{figChiAngle}. The solid line is a fit to the sinusoidal expression $H_{K,\theta} = H_K/\cos{\theta}$, with $H_K\simeq$ 16 T. Panel (b) plots the field profiles of the differential susceptibility $\chi_d$ at selected $T$ from 10 to 40 K measured in Sample 2 with $\theta$ fixed at 50$^\circ$. The lower $x$-axis displays the applied field $H$ while the upper $x$-axis displays the $z$ component $H_z$. At 40 K, $\chi_d$ is a featureless, decreasing function of $H$. Below 40 K, however, a broad peak becomes resolvable, with an amplitude that grows as $T\to$ 10 K. The peak in $\chi_d$ varies only weakly with $T$ from $H_z$ = 15 to 16 T (red arrows). 
}
\end{figure}

\begin{figure}[h]
\includegraphics[width=7cm]{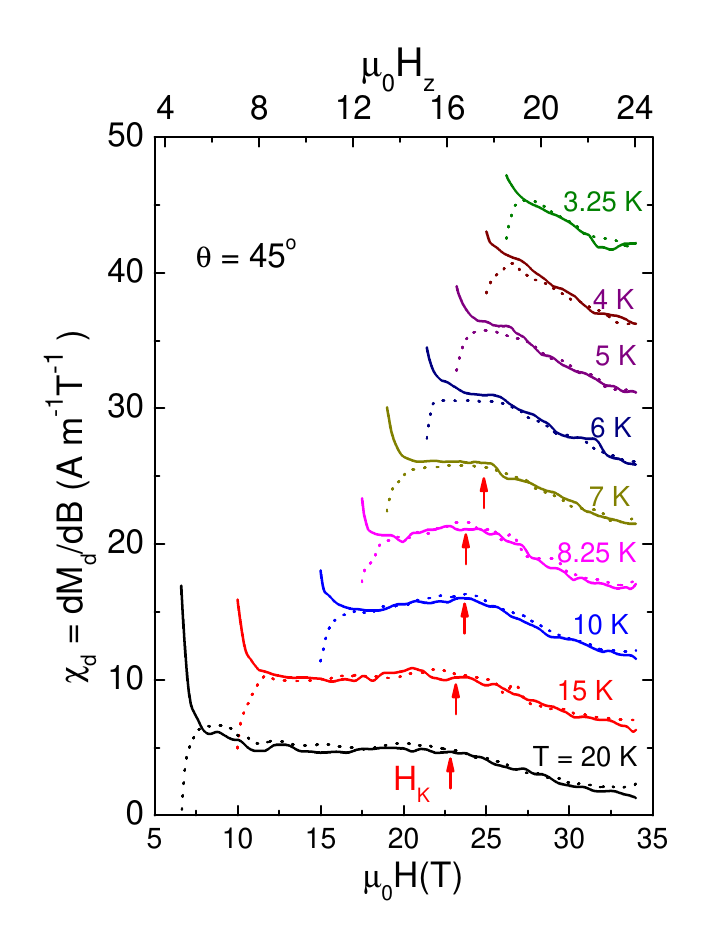}
\caption{\label{figdMHLowT} (color online) 
The differential susceptibility $\chi_d = \partial M/\partial B$ plotted vs. the applied field $H$ (lower $x$ axis) and $H_z = H\cos\theta$ (upper) measured in Sample 2 with $\theta = 45^\circ$ at selected temperatures 3.25 K $\leq T \leq$ 20K. The curves have been shifted vertically for clarity. At each $T$, the solid curves (dashed curves) were recorded as $H$ was slowly swept up (swept down). Below 6 K, the kink feature cannot be resolved against the large variations of hysteretic loops in the $\chi_d$-$H$ curves. }
\end{figure}

An important feature of the $M_d$-$H$ curves is the inflection (or kink) near $H$ = 25 T for $T$ below 20 K. 
The inflexion is observed as a weak maximum in the differential susceptibility $\chi_d= \frac{dM_d}{dB}$, with $B$ the flux density. Figure \ref{figChiAngle} shows the curves of $\chi_d$ vs. $H$ in measurements taken up to 45 T (Sample 3) at selected values of the field tilt-angle $\theta$ from 9$^\circ$ to 65$^\circ$ with $T$ fixed at 15 K. The peaks are observed in the reversible part of the vortex phase diagram (solid and dashed curves were recorded in the field sweep-up and -down directions, respectively). We refer to the maxima in $\chi_d$ at each $\theta$ as the field $H_{K,\theta}$ (identified by vertical arrows in Figs. \ref{figChiAngle}, \ref{figAngle}b, and \ref{figdMHLowT}). (In the curves of $\chi_d$ vs. $H$ plotted in Fig. \ref{figAngle}b, the broad peak sits on a gentle linear background. The background does not affect the actual value of $H_K$ which can be determined from the change in sign of $d\chi_d/dB$.)

The strong angular dependence of $H_{K,\theta}$ in Fig. \ref{figChiAngle} suggests that it is mostly determined by the $z$ component $H_z$ of the field. In Fig. \ref{figAngle}a we plot the variation of $H_{K,\theta}$ vs. $\theta$ at 15 K. As $\theta$ increases, $H_{K,\theta}$ increases monotonically from $\sim$15 T (at 10$^\circ$) to values exceeding our maximum applied field of 45 T when $\theta$ exceeds 65$^\circ$. By fitting the data to the sinusoidal form $H_{K,\theta} = H_K/\cos\theta$ (solid curve), we confirm that, for each $\theta$, the kink feature occurs when $H_z$ equals a $\theta$-independent field, which we call $H_K$. The peak signals an electronic transition that is driven predominantly by $H_z$. Below, we identify $H_K$ with the onset of static charge-ordering.

The field $H_K$ inferred from $\chi_d$ hardly varies at all with $T$ below 40 K. Figure \ref{figAngle}b plots the curves of $\chi_d$ vs. $H$ measured in Sample 2 at selected $T$ from 40 to 10 K. At 40 K, the curve of $\chi_d$ is initially featureless. As we cool to 10 K, a broad peak appears at 10 K. As shown in the scale on the upper $x$-axis, the peak (arrows) varies from 15 to 16 T. Figure \ref{figdMHLowT} plots curves of $\chi_d$ for $T$ down to 3.25 K in Sample 2. As we cool below 10 K, the rapid increase in melting field $H_m(T)$ causes the two fields to cross near 6 K, so that the $H_K$ feature now occurs within the vortex solid state where the variation of $M_d$ vs. $H$ is steep and hysteretic. This precludes determination of $H_K$ using the peak in $\chi_d$. However, as we show next, we can track it inside the vortex solid region using the thermal conductivity.

\begin{figure}[ht]
\includegraphics[width=7cm]{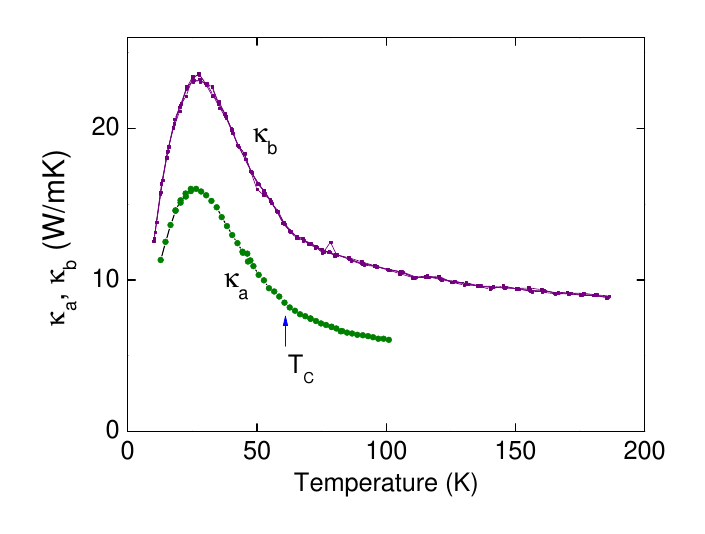}
\caption{\label{figKaKb} (color online) 
The thermal conductivities $\kappa_a$ (measured in Sample 5) and $\kappa_b$ (Sample 4) in zero $H$.
}
\end{figure}

\section{Thermal conductivity}
We have measured thermal conductivity in Samples 4 and 5 using a standard configuration with two thermometers and one resistive heater. The shorter edge of the rectangular sample platelet was rigidly attached to the heat bath with silver paint, standing upright in vacuum in an applied field $\bf H\parallel c$. In Sample 4, the heat current density $\textbf{J}_Q$ was applied parallel to the axis $\bf b$ (the chain axis), while in Sample 5 $\textbf{J}_Q \parallel a$ (Fig \ref{figKaKb}). To distinguish the two axes, we define the longitudinal thermal conductivities as $\kappa_b$ and $\kappa_a$, respectively ($\kappa_b > \kappa_a$), and suppress writing the tensor indices $xx$ and $yy$. Note that below $T_c$, for $H$ = 0, $\kappa_b$ exceeds $\kappa_a$ by a factor 1.3 to 1.4. In finite $H$, the factor is even larger at 4.5 K, as we show next.

\begin{figure}[ht]
\includegraphics[width=8cm]{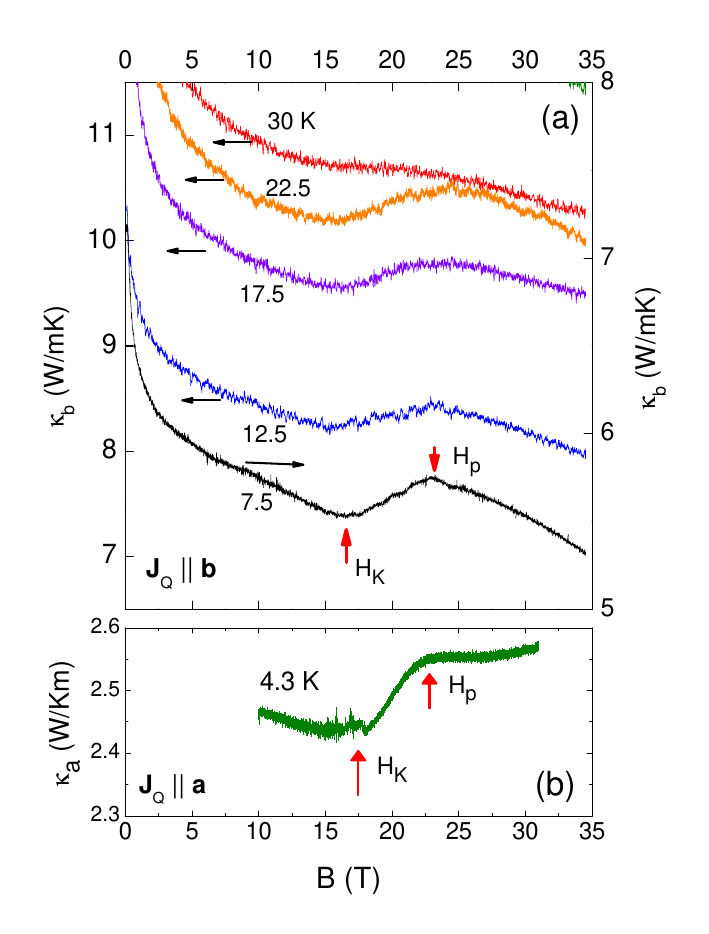}
\caption{\label{figKvsH} (color online) 
Expanded views of curves of the thermal conductivity at selected $T$. Panel (a) displays traces of $\kappa_b$ measured in Sample 4 measured in a field $\bf H \parallel c$ with the heat current density ${\bf J}_Q\parallel \bf b$ (chain axis). The vertical scale for 7.5 K is on the right axes. At the fields $H_K$ and $H_p$ (arrows), $\kappa_b$ exhibits a distinct changes of slope.
Panel (b) plots $\kappa_a$ measured at 4.3 K in Sample 5 with $\bf H\parallel c$ and ${\bf J}_Q\parallel \bf a$. 
}
\end{figure}

The heater was glued directly on the crystal using silver paint, while the chip thermometers (Lakeshore Cryotronics Cernox CX1030 bare chip) were heat sunk to the crystal (using epoxy) via thick gold wires. Four thin phosphor-bronze wires were attached to each thermometer for a four-point resistance measurement. This configuration ensures good thermal contact and fast equilibration between the thermometers and the sample. The thermometers were carefully calibrated \emph{in situ}, both in zero $H$ and large $H$. The low-temperature setup including two reference thermometers on the sample stage was contained in a custom-made vacuum can inside the variable-temperature insert in a 35-T resistive magnet (see SI Appendix, Sec. S4).

The curves of $\kappa_a$ and $\kappa_b$ vs. $H$, plotted in expanded scale in Fig. \ref{figKvsH}, reveal two distinctive features (breaks in slope) that are nearly $T$ independent below 40 K (arrows). The feature at lower $H$ may be identified with $H_K$ because it occurs at the same field value to our resolution. The upper feature is called $H_p$. As $T$ is raised, the features at $H_K$ and $H_p$ are thermally broadened, becoming unresolvable above 35 K. The step-changes in $\kappa_a$ and $\kappa_b$ in the field interval ($H_K$, $H_P$) are quite small, accounting for $< 5\%$ of the electronic thermal conductivity (see SI Appendix, Fig. S6).

\section{Magnetization to 45 Tesla and dHvA oscillations}
The foregoing experiments, carried out to maximum fields of 35 T, pointed to intriguing features in the magnetic phase diagram which become better resolved at lower $T$ and higher magnetic fields. We have extended the torque experiments to 45 T at temperatures down to 0.3 K and uncovered a distinct phase of the vortex system (vortex solid 2) that survives to 41 T. In addition, we have resolved de Haas van Alphen (dHvA) oscillations which onset in the vortex solid 2 phase (with amplitude and Dingle temperature closely similar to those in a previous dHvA experiment~\cite{Sebastian,SdHYBCO}). We show that the vortex liquid appearing above the melting field is robust to $H_z$ = 41 T (and well beyond judging from the trend vs. $H$).

\begin{figure}[ht]
\includegraphics[width=8cm]{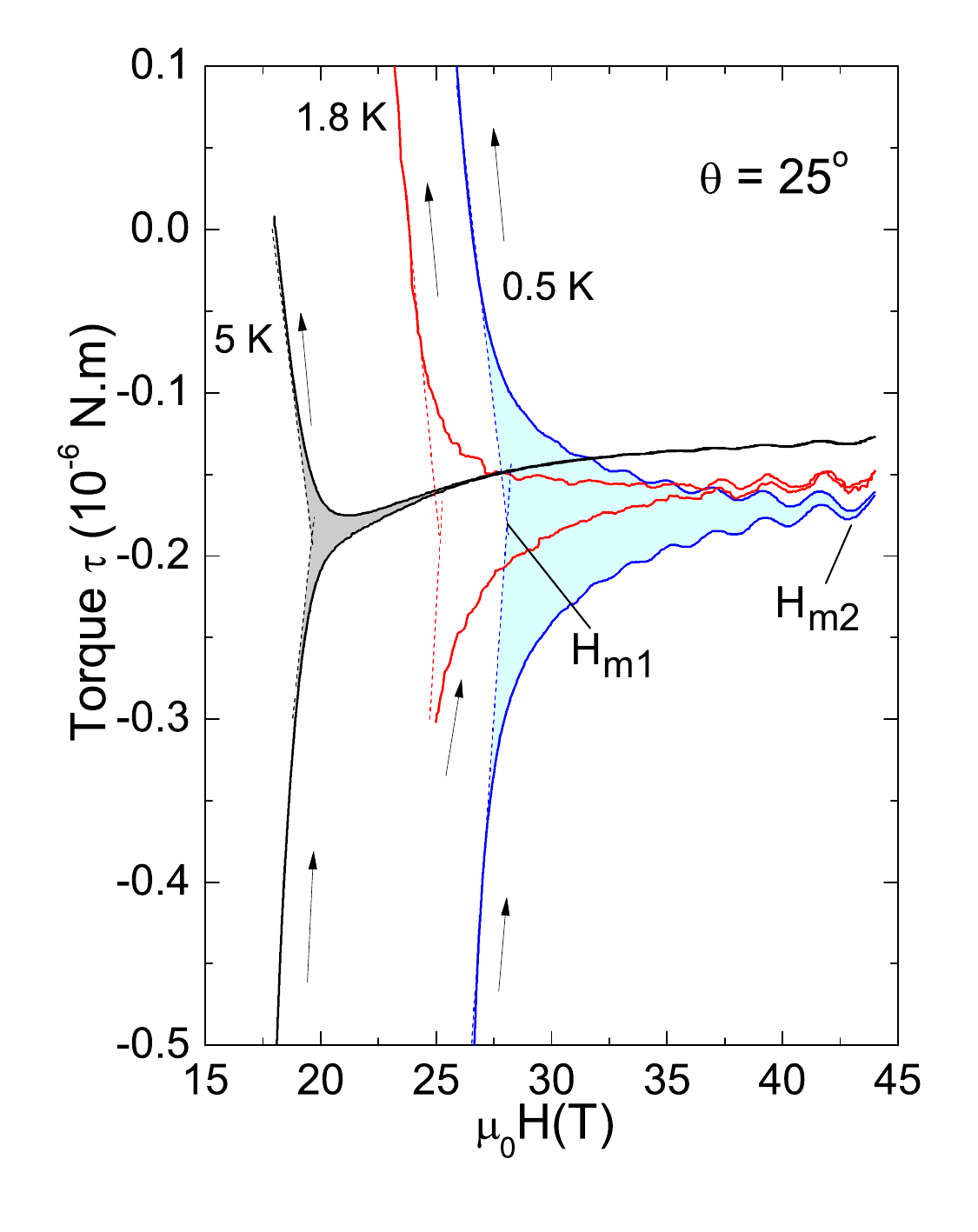}
\caption{\label{figtau} (color online) 
Hystereses in the curves of the torque $\tau$ vs. $H$ in high fields at low $T$ in Sample 3. At each $T$, hysteretic behavior between the sweep-up and -down curves defines the field region in which the vortex solid is stable (arrows indicate field sweep directions). For the curves at 15 K (black curves), the rapid closing of the hysteresis loop with increasing $H$ (gap between the upper and lower branches) signals the melting of the vortex solid. The nexus of the linear extrapolations of the two branches (dashed lines) defines the melting field $H_{m1}\sim$19.5 T. However, a small wedge representing vortex solid 2 exists to $H_{m2}\sim$ 25 T (shaded grey). At 0.5 K, the vortex solid 2 region has expanded considerably (shaded light blue). At the 3 temperatures, dHvA oscillations are resolved. At 0.5 K, they onset in the vortex solid 2 region. We note that, at all fields up to 45 T, the magnetization is manifestly strongly diamagnetic ($\tau <0$). Hence, at each $T$, the region above $H_{m2}$ is a stable vortex liquid with finite pair amplitude. 
}
\end{figure}

Figure \ref{figtau} shows traces of the torque $\tau$ vs. $H$ in the field region where the hysteretic loops begin to close at $T$ = 0.5, 1.8 and 5 K. A number of important conclusions may be drawn from Fig. \ref{figtau}. 
First, the rapid closing of the hysteresis loop suggests that we can define a ``lower'' melting field $H_{m1}$ (the nexus of the linear extrapolations of the upper and lower branches of $\tau$). Above $H_{m1}$, there exists a second vortex solid phase (``vortex solid 2'', in the shaded region) that has a much weaker shear modulus but survives to a higher melting field $H_{m2}$. As we cool from 5 to 0.5 K, the small shaded region expands rapidly ($H_{m1}$ = 28 T and $H_{m2}\sim$ 42 T at 0.5 K). The rapid growth suggests that the vortex solid 2 constitutes a distinctive, stable phase of the overall vortex solid in the limit $T\to 0$ (instead of a fluctuation tail).

\begin{figure}[h]
\center{\includegraphics[width=9cm]{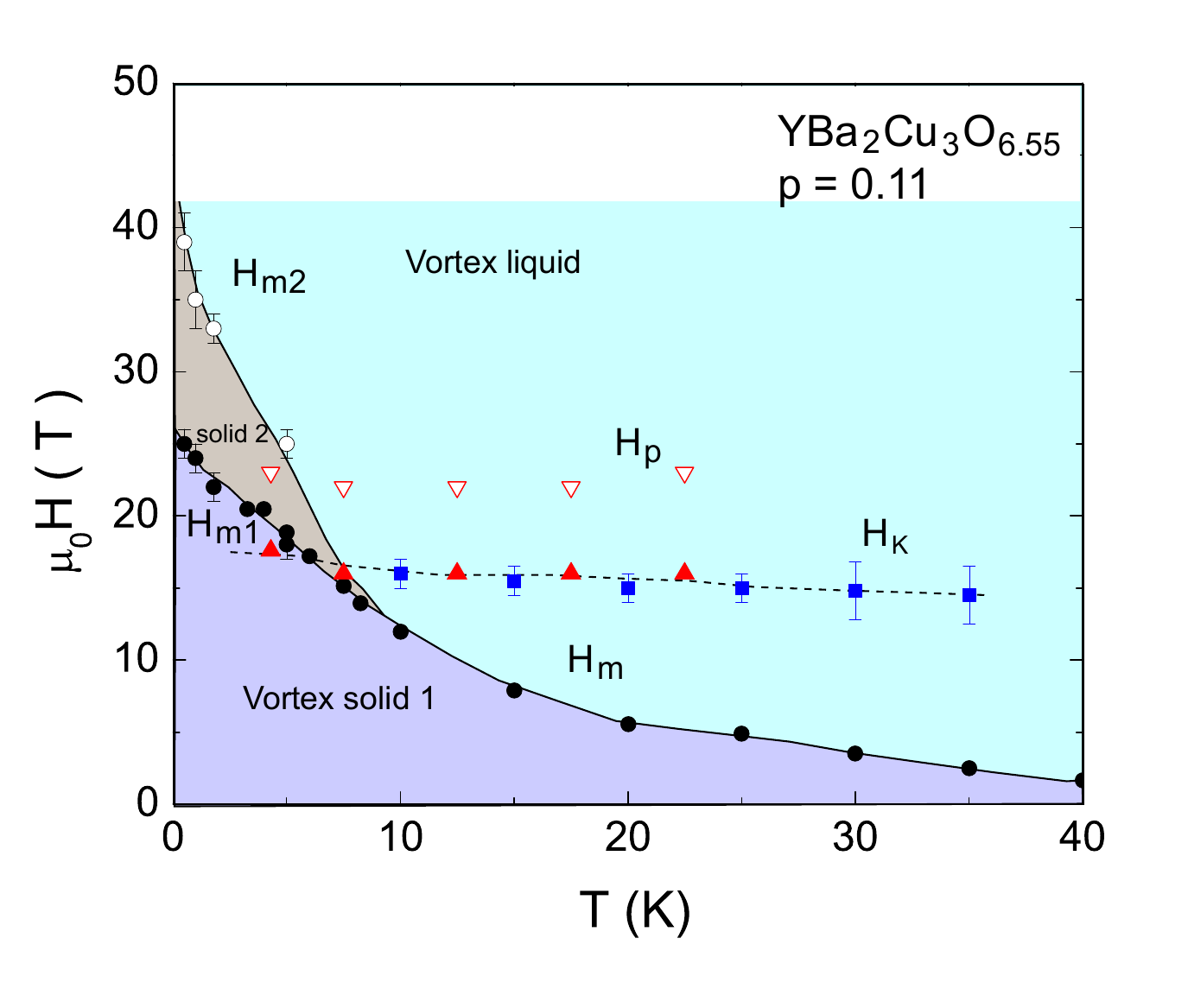}}
\caption{\label{figphase} (color online) 
The magnetic phase diagram in untwinned YBa$_2$Cu$_3$O$_y$ ($y$ = 6.55) inferred from magnetization and the thermal conductivity.  Given the 2D nature of the transition curves, the vertical axis refers to the $z$-component of $\bf H$ applied in the torque experiments (for $\kappa_{xx}$ measurements, $\bf H$ is always $\parallel \hat{z}$). Above $\sim$8 K, the vortex solid is stable below the melting field $H_m$ (solid circles). Below $\sim$8 K, the $H_m$ curve splits into two branches $H_{m1}$ (solid circles) and $H_{m2}$ (open circles). The vortex solid 1 below $H_{m1}$ displays a large critical current density $J_c$ and shear modulus, whereas the solid 2 ($H_{m1} < H < H_{m2}$) has a much smaller $J_c$ but survives to $\sim$41 T at 0.5 K. Throughout, the vortex-liquid state is stable to at least 41 T (region shaded light blue), but likely much higher judging from the trend in $M_d$ vs. $H$. The nearly $T$ independent kink field $H_K$, identified as the onset of static charge order, is plotted as blue squares (inferred from $\chi_d$) and as red solid triangles (from $\kappa_a$ and $\kappa_b$). It intersects $H_{m1}$ without affecting the melting (to our resolution). The field scale $H_p$ (open triangles, derived from $\kappa_a$ and $\kappa_b$) appears to terminate at the lower melting field $H_{m1}$ in the limit $T\to 0$. At 0.5 K, the dHvA oscillations onset near 25 T (coexist with the vortex liquid below $\sim$41 T).
}
\end{figure}

Secondly, when the curves are non-hysteretic ($H>H_{m2}$), the magnetic response is strongly diamagnetic at all fields up to 45 T ($\tau < 0$). From the trends of the curves, the diamagnetic response extends to fields considerably above 45 T. Because reversible diamagnetism is a hallmark of the vortex liquid, we conclude that the vortex liquid and the underlying pair amplitude are stable to extremely high magnetic fields, well beyond either melting field. These observations confirm the conclusion from previous Nernst and torque experiments that the pairing strength in cuprates is extremely robust.

Finally, dHvA oscillations can be resolved even in the curves at 5 K (see SI Appendix, Sec. S3, Fig. S4). The onset of the oscillations at 0.5 K occurs within the shaded region. From the discussion above, we conclude that the pairing amplitude is non-zero when the oscillations onset. Above $H_{m2}$, the oscillations exist in the vortex liquid state where the magnetization is manifestly diamagnetic. The coexistence of the dHvA oscillations with diamagnetic response was previously observed in Ref.~\cite{Sebastian}.

\section{Magnetic Phase Diagram and Discussion}
The combined experiments uncover a rather rich magnetic phase diagram that allows us to relate the charge-ordering field scales obtained by NMR and x-ray diffraction to the vortex states in underdoped YBCO. In Fig. \ref{figphase} the melting field $H_m$, which defines the stability region of the vortex solid, increases monotonically as $T$ decreases below $T_c$ (black circles). At $\sim 9$ K, it bifurcates into the two curves $H_{m1}$ and $H_{m2}$. Whereas $H_{m1}$ intercepts the $H$ axis at $\sim 25$ T at 0.5 K (our lowest $T$), the higher field $H_{m2}$ rises steeply to exceed 40 T at 0.5 K. The two vortex solid phases separated by $H_{m1}$ are shaded differently and labelled as 1 and 2. As discussed above, $H_{m1}$ -- estimated by a linear extrapolation of the sweep-up and sweep-down curves of $M_d$ vs. $H$ -- measures the field at which the vortex solid loses most of its shear rigidity. Above $H_{m1}$, however, substantial irreversibility remains which implies that the vortex solid remains. This is most apparent in the $M_d$-$H$ curves at 0.5 K (Fig. \ref{figtau}).

One of our key conclusions is that the vortex liquid is stable in the region shaded light blue in Fig. \ref{figphase}. This region, extending from $H_m$ to our maximum $H_z\sim$ 41 T and up to temperatures above $T_c$. We emphasize that, even when hysteretic behavior vanishes in the $M_d$-$H$ curves, the magnetization remains negative and strongly $T$ dependent (see especially the curve at 0.5 K in Fig. \ref{figtau}). The reversible, $T$-dependent diamagnetism is the experimental hallmark of the vortex liquid. The pair amplitude is non-zero in the vortex liquid although dissipation is large because the vortex liquid -- lacking a finite shear modulus -- cannot be pinned by disorder. The picture of a relatively small $H_m(T)$, together with the very large pairing energy scale that allows the vortex liquid to survive to very large $H$, is closely similar to that in other underdoped cuprates~\cite{Wang2005,Wang2006,Li2007,Li2010,Kivelson,Bernhard2011}.

In Fig. \ref{figphase}, we have also plotted the field $H_K(T)$ inferred from $\chi_d$ (blue squares) with the dashed line drawn as guide. The red triangles represent $H_K$ inferred from $\kappa_{xx}$. As $T$ decreases from 35 to 4 K, $H_K$ is nearly $T$ independent. Below 6 K, $H_K$ intersects the melting curve. We identify $H_K$ with the onset of LRO in the CDW as detected by x-ray diffraction (15 T at 10 K)~\cite{Gerber}. 3D ordering of the CDW onsets near 15 T although strong 2D fluctuations appear at 10 T~\cite{Chang16}. In Wu \emph{et al.}~\cite{TaoWu2013}, the splitting of the NMR $^{17}O$ lines onsets at a lower field (10 T for $p$ = 0.109), but saturates at a field close to our $H_p$ (23 T). Below 40 K, the field scale $B_{co}$ extracted from an ultrasonic experiment~\cite{LeBoeuf} lies slightly higher than our $H_K$ curve (although the difference may arise from the combined uncertainties). The present results imply that the initial onset of static charge ordering has no observable effect on the melting field of the vortex solid.

The higher field $H_p$ inferred from $\kappa_{xx}$ (also nearly $T$ independent) is plotted as open triangles. In the limit $T\to 0$, it extrapolates to a value $24\pm 1$ T close to the intercept of $H_{m1}$. $H_p$ is close to the value at which the splitting of the NMR lines saturates. 

A recent experiment in UD YBCO (with $p$ = 0.11) has observed that $\kappa_{xx}$ at 1.8 K rises to a distinct peak at $\sim$23 T~\cite{Grissonnanche}. The authors interpret the peak as the true upper critical field $H_{c2}$. However, the present results show that, at $T$ = 0.5 K, the vortex solid exists up to 41 T while the vortex liquid persists to even higher $H$. Hence the pair condensate with strong superconducting correlations survives well above 23 T. A similar inference based on the x-ray peak intensity at 28 T was reported by Gerber \emph{et al.}~\cite{Gerber}.

The different experimental probes provide a picture that is richer and more interesting than the identification~\cite{Grissonnanche} of a mean-field BCS-type transition at which the pair condensate simply vanishes (at 23 T). In a $d$-wave superconductor, the electronic thermal conductivity $\kappa^e$ (we suppress the subscript $xx$) is overwhelmingly dominated by the quasiparticle (qp) population at the nodes (at $T\ll T_c$). Hence changes to $\kappa^e$ predominantly reflect changes to the FS at or close to the node (example, closing of the gap along a FS arc or creation of a small pocket). Moreover, the curves in Fig. \ref{figKvsH} show that raising $T$ above 25 K suppresses all traces of the step-change in $\kappa_{xx}$ (5$\%$ at 4.3 K). Thermal broadening of the broad peak at $H_K$ in $\chi_d$ is equally rapid (Fig.\ref{figAngle}). This provides direct evidence that the FS changes at the node involve energy scales of $\sim$1-2 meV, far smaller than the gap amplitude $\Delta_0$ at the antinodes. In Fig. \ref{figphase}, the convergence of the lower melting curve $H_{m1}$ (which separates the two vortex solids) and the crossover curve $H_p$ to values close to 23 T as $T\to 0$ suggests to us that, as the 3D CDW amplitude grows, the pair condensate undergoes an abrupt transition at 23-24 T to accommodate the competing CDW (this strongly affects the vortex solid shear modulus as discussed below). We propose that the peak in $\kappa^e$ reflects changes (at energy scales of 1-2 meV) to the nodal states caused by the CDW transition, rather than suppression of the gap amplitude $\Delta_0$ (40-80 meV).

In Ref. \cite{Grissonnanche}, a mean-field $H_{c2}$ curve extending from 23 T at 2 K to zero at $T_c$ is drawn without reference to experiment. The magnetization curves in Fig. \ref{figMH} are incompatible with such a mean-field curve ($M_d$ varies smoothly through the proposed $H_{c2}$ curve; see arrow labelled MF). [After the present manuscript was submitted, we extended measurements of both $\kappa_{xx}$ and the thermal Hall conductivity $\kappa_{xy}$ to lower $T$. For the doping level in our crystals, we observe a sharp cusp similar to that in Ref.~\cite{Grissonnanche} at 18 T at 0.5 K (lower than 23 T, but consistent with the dashed line linking solid triangles in Fig. \ref{figphase}). As 18 T lies below $H_{m1}$, the system is well within the vortex solid 1 phase when the transition affecting the nodal qp states occurs.]

An interesting issue raised by the phase diagram is the fate of both the vortex solid and liquid in the quantum limit $T\to 0$~\cite{Li2007}. As remarked above (see Fig. \ref{figtau}), the rapid expansion of the vortex solid 2 phase (between $H_{m1}$ and $H_{m2}$) suggests that the pair condensate may modify its pairing pattern to accomodate the static CDW. Several groups have proposed~\cite{Tranquada,Ogata,Vafek,PALee,Fradkin,Hamidian} either the pair density wave (PDW) formation or the related Fulde Ferrell Larkin Ovchinikov (FFLO) state~\cite{Fulde}. It is intriguing that, when $T\to 0$, the transition at $H_{m1}\sim$ 24 T also occurs close to the field $H_p$. The present experiments show that this allows the vortex solid to survive to much higher $H$ at the cost of a substantial decrease in its shear modulus.

The question whether the vortex liquid exists as a stable quantum phase at $T = 0$ is resolvable by extending the torque measurements below 0.1 K to fields above 41 T. The properties of the quantum vortex liquid (in an ultraclean superconductor) are largely unknown but may become accessible to torque magnetometry at mK temperatures in the near future. The existence of the vortex liquid to fields substantially higher than 41 T implies that the Cooper pair amplitude in underdoped YBCO is too strong to be suppressed at these field scales.

\section{Acknowledgments}
The research  of L.L. is supported by the U.S. Department of Energy, Office of Basic Energy Sciences, Division of Materials Sciences and Engineering under Award DE-SC0008110 (high field magnetization). M.H. and N.P.O. were supported by NSF-MRSEC grant DMR 1420541 and the Gordon and Betty Moore Foundation's EPiQS Initiative through Grant GBMF4539 (thermal conductivity and analysis). The experiments were performed at the National High Magnetic Field Laboratory, which is supported by National Science Foundation Cooperative Agreement No. DMR-1157490, the State of Florida, and the U.S. Department of Energy.


\end{document}